\documentclass[fleqn,usenatbib]{mnras}

\usepackage{newtxtext,newtxmath}

\usepackage[T1]{fontenc}

\DeclareRobustCommand{\VAN}[3]{#2}
\let\VANthebibliography\thebibliography
\def\thebibliography{\DeclareRobustCommand{\VAN}[3]{##3}\VANthebibliography}

\usepackage{graphicx}	
\usepackage{amsmath}	

\def\citejap#1{\citeauthor{#1}\ \citeyear{#1}}

\title[Gravitational redshifting of galaxies in SPIDERS]{Gravitational redshifting of galaxies in the SPIDERS cluster catalogue}

\author[C. T. Mpetha et al.]{C. T. Mpetha$^{1,2}$,\thanks{E-mail: c.mpetha@ed.ac.uk}
C. A. Collins$^{1}$,
N. Clerc$^{3}$,
A. Finoguenov$^{4}$, 
J. A. Peacock$^{2}$,
J. Comparat$^{5}$,\
\newauthor D. Schneider$^{6}$, 
R. Capasso$^{7}$,
S. Damsted$^{4}$,
K. Furnell$^{1}$,
A. Merloni$^{5}$,
N. D. Padilla$^{8}$,
A. Saro$^{9,10,11,12}$
\\
$^{1}$Astrophysics Research Institute, Liverpool John Moores University, IC2 Building, Liverpool Science Park, 146 Brownlow Hill, Liverpool  L3 5RF\\
$^{2}$Institute for Astronomy, University of Edinburgh,
Royal Observatory, Blackford Hill, Edinburgh, EH9 3HJ, United Kingdom\\
$^{3}$IRAP, Universite de Toulouse, CNRS, UPS, CNES, F-31028 Toulouse, France\\
$^{4}$Department of Physics, University of Helsinki, PO Box 64, FI-00014 Helsinki, Finland\\
$^{5}$ Max-Planck-Institut f{\"u}r extraterrestrische Physik (MPE), Giessenbachstrasse 1, D-85748 Garching bei M{\"u}nchen, Germany \\
$^{6}$Institute for Gravitation and the Cosmos, Pennsylvania State University, University Park, PA 16802, USA \\
$^{7}$ The Oskar Klein Centre, Department of Physics, Stockholm University, AlbaNova University Center, SE-106 91 Stockholm, Sweden\\
$^{8}$Instituto de Astrofisica, Universidad Catolica de Chile, V. Mackenna 4860, Santiago, Chile\\
$^{9}$Astronomy Unit, Department of Physics, University of Trieste, via Tiepolo 11, I-34131 Trieste, Italy\\
$^{10}$IFPU - Institute for Fundamental Physics of the Universe, Via Beirut 2, 34014 Trieste, Italy\\
$^{11}$INAF - Osservatorio Astronomico di Trieste, via G. B. Tiepolo 11, I-34143 Trieste, Italy\\
$^{12}$INFN - National Institute for Nuclear Physics, Via Valerio 2, I-34127 Trieste, Italy\\
}

\pubyear{2020}

\begin{document}
\label{firstpage}
\pagerange{\pageref{firstpage}--\pageref{lastpage}}
\maketitle

\begin{abstract}
 Data from the SPectroscopic IDentification of ERosita Sources (SPIDERS) are searched for a detection of the gravitational redshifting of light from $\sim\!20\,000$ galaxies in $\sim\!2500$ galaxy clusters using three definitions of the cluster centre: its Brightest Cluster Galaxy (BCG), the redMaPPer identified Central Galaxy (CG), or the peak of X-ray emission. Distributions of velocity offsets between galaxies and their host cluster's centre, found using observed redshifts, are created.  The quantity $\hat{\Delta}$, the average of the radial velocity difference between the cluster members and the cluster systemic velocity, reveals information on the size of a combination of effects on the observed redshift, dominated by gravitational redshifting. The change of $\hat{\Delta}$ with radial distance is predicted for SPIDERS galaxies in General Relativity (GR), and $f(R)$ gravity, and compared to the observations. The values of $\hat{\Delta}=-13.5\pm4.7$\,km\,s$^{-1}$, $\hat{\Delta}=-12.5\pm5.1$\,km\,s$^{-1}$, and $\hat{\Delta}=-18.6\pm4.8$\,km\,s$^{-1}$ for the BCG, X-ray and CG cases respectively broadly agree with the literature. There is no significant preference of one gravity theory over another, but all cases give a clear detection ($>2.5\sigma$) of $\hat{\Delta}$. The BCG centroid is deemed to be the most robust method in this analysis, due to no well defined central redshift when using an X-ray centroid, and CGs identified by redMaPPer with no associated spectroscopic redshift. For future gravitational redshift studies, an order of magnitude more galaxies, $\sim\!500\,000$, will be required—a possible feat with the forthcoming \textit{Vera C. Rubin Observatory}, \textit{Euclid} and \textit{eROSITA}.
\end{abstract}




\section{Introduction}


Galaxy clusters are the largest gravitationally bound systems in the Universe, making them an excellent test-bed for theories of gravity. They are composed of $\sim\!10^{2}-10^{3}$ galaxies and a large dark matter halo.
\begin{equation}
    \ln(1 + z_{\rm obs}) = \ln(1+z_{\rm cos})+\ln(1+z_{\rm pec})+\ln(1+z_{\rm grav}) \, .
    \label{eq:zcont}
\end{equation}
 There are various effects that contribute to the observed redshifting of light from galaxies in clusters ($z_{\rm obs}$), shown in equation (\ref{eq:zcont}). There is of course the cosmological redshift ($z_{\rm cos}$) due to the expansion of the Universe, which will be the same for both the galaxy and the host galaxy cluster. After this, the most prominent is the peculiar redshift ($z_{\rm pec}$)—random isotropic motions of galaxies within the cluster in the line of sight. Galaxies are in motion around the minimum of the cluster's potential well, its dynamical centre, and so the average offset between a galaxy's peculiar redshift and that of the cluster centre will be zero. To test this, a distribution of line-of-sight velocity offsets, found from observed redshifts, can be created. If the peculiar redshift were the only contribution along with the cosmological redshift, this distribution would be centred on zero, due to isotropy. But this is not the case, and so the shift of the centre of this distribution is informative of the size of other contributions, namely gravitational redshifting ($z_{\rm grav}$) whose possibility of detection was investigated by \citet{Nottale} and \citet{Cappi}. This shifting of the average is  the quantity of interest in this study; the size of the shift and its evolution with distance from the cluster centre are both informative on the theory of gravity governing the observed redshifts of these galaxies. To create a distribution of \textit{line-of-sight} velocity offsets using galaxy redshifts, we define the quantity
\begin{equation}
    \Delta = c\left[\ln(1 + z_{\rm obs}) -\ln(1+z_{\rm cen})\right] \, ,
\end{equation}
where $z_{\rm cen}$ is the redshift of the galaxy cluster's centre. Differences in the logarithm of the redshifts have been used instead of simply assuming $z = v/c$ as using the natural logarithm provides a better approximation to the line-of-sight velocity \citep{zeta}:
\begin{equation}
    \ln(1+z) \simeq  \frac{v_{\rm los}}{c} + \frac{1}{2}\left(\frac{v^{2}}{c^{2}}-\frac{v_{\rm los}^{2}}{c^{2}}\right) + \cdots \, .
\end{equation}
Another advantage of this definition is that it removes some of the dependence on the cosmological redshift, as otherwise the expression would be $\Delta = c(z_{\rm obs}-z_{\rm cen})/(1+z_{\rm cen})$. This is advantageous if the cosmological redshift has a large uncertainty.

$\Delta$ is a combination of the peculiar velocity, gravitational redshift and other effects that will be detailed in section \ref{sec:delta}. Then, to find the location of a distribution of $\Delta$ values, we define
\begin{equation}
    \hat{\Delta} \equiv\left\langle \Delta \right\rangle \, .
\end{equation}
Following \cite{BWT}, Tukey's biweight average is used as a minimum variance estimator for the location and scale of galaxy velocity distributions, which are in general not Gaussian due to the presence of interlopers, and dynamical instability of the cluster.

This paper will use those SDSS Data Run 16 galaxies and clusters that have been spectroscopically measured as part of the SPIDERS programme  \citep{Clerc}, to explore the size of $\hat{\Delta}$ at different distances from the centre of a cluster. \citet{WHH} made the first tentative detection of gravitational redshifting in galaxy clusters, with data from SDSS Data Run 7 \citep{SDSS7}. A similar analysis has been performed on SDSS DR10 galaxies and clusters \citep{GRSDSSBOSS}. This paper aims to repeat and build upon these analyses. The large amount of information available from the spectroscopic follow up of X-ray selected galaxy clusters allows for novel methods of calculating $\hat{\Delta}$ as a function of distance from the cluster centre. 

The outline of this paper is as follows; in section \ref{sec:data} the SPIDERS catalogue is introduced and the properties of its clusters are discussed, including three different definitions of the centre of a cluster, then the data reduction process is presented. Next in section \ref{sec:delta} the various contributions to $\hat{\Delta}$ are described, before its variation with distance from the centre of the cluster is predicted for two theories of gravity in section \ref{sec:predict}. In section \ref{sec:res} $\hat{\Delta}$ is found for each of the three centroid cases from observations of galaxies and clusters in the SPIDERS catalogue, and compared to the predictions. We conclude with closing remarks and future prospects in section \ref{sec:conc}.

\citet{planck} values of $H_{0} = 67.4$\,km\,s$^{-1}$\,Mpc$^{-1}$ and $\Omega_{\rm m,0}=0.315$ have been used throughout.

\section{data}
\label{sec:data}
\subsection{SPIDERS catalogue}

SPectroscopic IDentification of \textit{eROSITA} Sources (SPIDERS: \citejap{Clerc}, \citeyear{Clerc2}; \citejap{SPIDERSconst}; Kirkpatrick et al. in prep.) is the X-ray specific subprogramme of the extended Baryonic Oscillation Spectroscopic Survey (eBoss: \citejap{eBOSS}), which is a part of the Sloan Digital Sky Survey (SDSS: \citejap{SDSS}; \citejap{SDSS2}). The SDSS is currently in its fourth generation, SDSS-IV \citep{SDSSIV}. SPIDERS is the spectroscopic follow up of large numbers of galaxies identified in the eBOSS survey. Galaxies are assigned to clusters using the redMaPPer algorithm \citep{RM}, which then uses identified members to estimate cluster properties such as optical richness and redshift.

SPIDERS will eventually include \textit{eROSITA} X-ray selected clusters. The most recent catalogue, Data Run 16, comprises a subset of clusters that were identified in the CODEX program \citep{CODEX}, which searched ROSAT All Sky Survey data for extended X-ray sources. Data Run 16 contains 2740 clusters with close to 42\,000 galaxies with spectroscopic redshifts. There are a number of parameters measured for each cluster, including the virial mass $M_{200}$ estimated from its velocity dispersion, and both $M_{\rm 200}$ and the X-ray luminosity $L_{\rm X}$ are iteratively calibrated using an $M_{\rm 200}-L_{\rm X}$ scaling relation \citep{CSCALE2}. Also provided within SPIDERS are three potential definitions of the centre of a cluster, each of which has been used for an independent measurement of $\hat{\Delta}$.
\subsection{Cluster properties}

There are numerous reasons why it is advantageous to have a large cluster sample. Firstly, even assuming that every galaxy in a cluster could be spectroscopically measured, there are simply too few galaxies to allow the statistical detection of a non-zero gravitational redshift. Typically $\hat{\Delta} \sim10$\,km\,s$^{-1}$, and a cluster's velocity dispersion is $\sigma_{\rm v} \sim 1000$\,km\,s$^{-1}$. To have a standard error on the average value  of the distribution, $\sigma_{\rm v}/\sqrt{N}$, that is small enough to resolve the gravitational redshifting from no effect, around $10\,000$ galaxies are required. By stacking many galaxy measurements from many clusters into a composite cluster, this requirement can be satisfied. Secondly, clusters do not in general exhibit spherical symmetry. There is often apshericity in the matter distribution leading to anisotropic velocity distributions. By stacking a large number of clusters, these features will be smoothed out in the composite cluster.

When stacking these clusters, simply using a distance in Mpc is not ideal, as clusters can have a large range of sizes, and so have different masses and densities at the same distance from the centre. Clusters show a high degree of similarity in their virialised region \citep[e.g.][]{similar}. For this reason the ratio $\tilde{r} = r/r_{\rm 200}$ is used as a distance measure. The virial radius $r_{200}$ is the radius of the cluster within which the mean density is equal to the overdensity parameter $v=200$ multiplied by the critical density of the Universe $\rho_{\rm c}$; the density of a flat Friedmann-Lemaître-Robertson-Walker Universe, at the redshift of the cluster. Throughout this paper it is assumed that clusters follow the Navarro-Frenk-White (NFW) density profile \citep{NFW}; a similar density profile across all clusters is thus assumed at similar values of $\tilde{r}$, and hence the effects in each cluster can be stacked and compared. In SPIDERS, the size of a cluster's virial radius on the sky is measured in degrees. Hence to find the distance of a galaxy from its parent cluster's centre in units of the virial radius, the ratio $\Delta\theta/\theta_{200}$ is calculated:\begin{equation}
    \tilde{R} = \frac{\Delta\theta}{\theta_{200}} \, .
    \label{eq:thetaR}
\end{equation}
$\tilde{R}$ describes the \textit{projected} distance of the galaxy from the cluster centre. By only knowing angular positions, information on the true radial distance is lost. This is fine, so long as when calculating the size of $\hat{\Delta}$ as a function of distance in different theories of gravity, it is done using the projected distance from the cluster centre.

\subsection{Defining the cluster central position and redshift}
\label{sec:cluscen}
Cluster miscentring is a leading cause of systematic error in cluster velocity dispersion analyses \citep{miscen}, warranting careful discussion of how the centre has been defined, and comparing possible methods. Within the SPIDERS catalogue are three methods of defining the cluster centre, and each one has been used for an independent measurement of $\hat{\Delta}$ in SPIDERS clusters.

The \textbf{Brightest Cluster Galaxy} (BCG) is the most luminous galaxy in the cluster. This can be used as a proxy for the centre of a cluster, as it is formed through the merger of other large galaxies—and so is likely to trace the dynamical centre, which would lie at the minimum of the clusters substantial potential well \citep{BCG}. The redshift and central position of the cluster is then taken to be the redshift and position of the BCG. In previous $\hat{\Delta}$ analyses this is the method by which the cluster centre is defined.

The \textbf{Optical Centre} definition is found from the red-sequence Matched-filter Probabilistic Percolation (redMaPPer) algorithm, which uses a probabilistic approach to find the most likely Central Galaxy (CG) \citep{RM}. It is assumed each cluster has a single dominant galaxy at its centre, which is a red sequence galaxy. Potential CGs are assigned a centre probability based on three observables; their $i-$band magnitude, red sequence photometric redshift, and the cluster density around the candidate CG. The most likely CG is identified, and thus used as the central position and redshift of the cluster. In many cases this coincides with the BCG, but there are enough differences for it to be an independent method. 

The \textbf{X-ray centre} is found using the peak of X-ray emission. This poses a problem of how the redshift of the centre is defined. One potential way of addressing this is by isolating the core of the cluster, centred on the X-ray peak, identifying all the galaxies lying in this region, and using their average as a measure of the central redshift.
Typical core radii are in the range $r_{\rm c} \simeq (0.1-0.25)h^{-1}$\,Mpc  \citep{coresize}. As the typical cluster virial radius in the SPIDERS catalogue is $r_{200} \simeq 1.5$\,Mpc , this gives a range of core radii of $r_{c} \simeq (0.095-0.24) r_{200}$.  To give the best chance of observing sufficient galaxies in the core region for a reasonable average, without diluting the $\hat{\Delta}$ signal by using a core radius too large for the majority of clusters, a core radius of\begin{equation}
    r_{\rm \rm c} = 0.2 r_{200}
\end{equation}
is henceforth used. Galaxies within this region are used to find the redshift of the X-ray defined centre. There is an immediate problem with this method: only information on the \textit{projected} distance from the cluster centre is known, so it is likely that in some cases galaxies not in the core region are being used to estimate the central redshift. Furthermore, the ROSAT centroid for very faint sources is poorly determined, and so there is a strong possibility of miscentring in these clusters.

Each method for defining the centre of the cluster has advantages and disadvantages. For the BCG and optical centroid cases, the effects contributing to $\hat{\Delta}$ depend on the motion of the CGs themselves, which are not at rest relative to the clusters potential minimum, and even the increased internal dynamics of these large galaxies can have an impact. These effects cause various slight adjustments to the values shown in Fig. \ref{fig:GR}. However in \citet{Kaiser} the net effect of these various modifications due to CG properties is found to be small ($\lesssim 1$\,km\,s$^{-1}$), and only affecting the innermost region where the CGs lie, and so for brevity they have been neglected in this analysis. Regardless, numerous studies \citep[e.g.][]{BCGvsXRAY} have found the BCG correlates well with the minimum of the gravitational potential. Using an X-ray centroid can be preferable as it avoids miscentring on foreground/background galaxies, and represents a better tracer in highly dynamical clusters. The obvious drawback in this case is the lack of a clear central redshift. In the optical centroid case, the redMaPPer algorithm is not perfect. It requires the central galaxy to be a red-sequence member, and so fails when the CG is undergoing strong star formation \citep{RM}. Another problem is that for SPIDERS clusters there is a large discrepancy ($\sim\!0.1r_{200}-0.3r_{200}$) between the optical centre and the nearest spectroscopically observed galaxy in 188 clusters, with around 400 clusters showing a largest difference between $0.002r_{200}$ and $0.1r_{200}$. This would suggest that in the more extreme of these cases redMaPPer has identified a CG that is not spectroscopically measured, while the smaller offsets are likely to be down to positional inaccuracies. To ensure there is no accidental miscentring, only galaxies within $0.033 r_{200} \sim 50$\,kpc, about the size of a large galaxy, of the optical centre have been identified as a CG.

\begin{figure}
    \centering
    \includegraphics[width=\columnwidth]{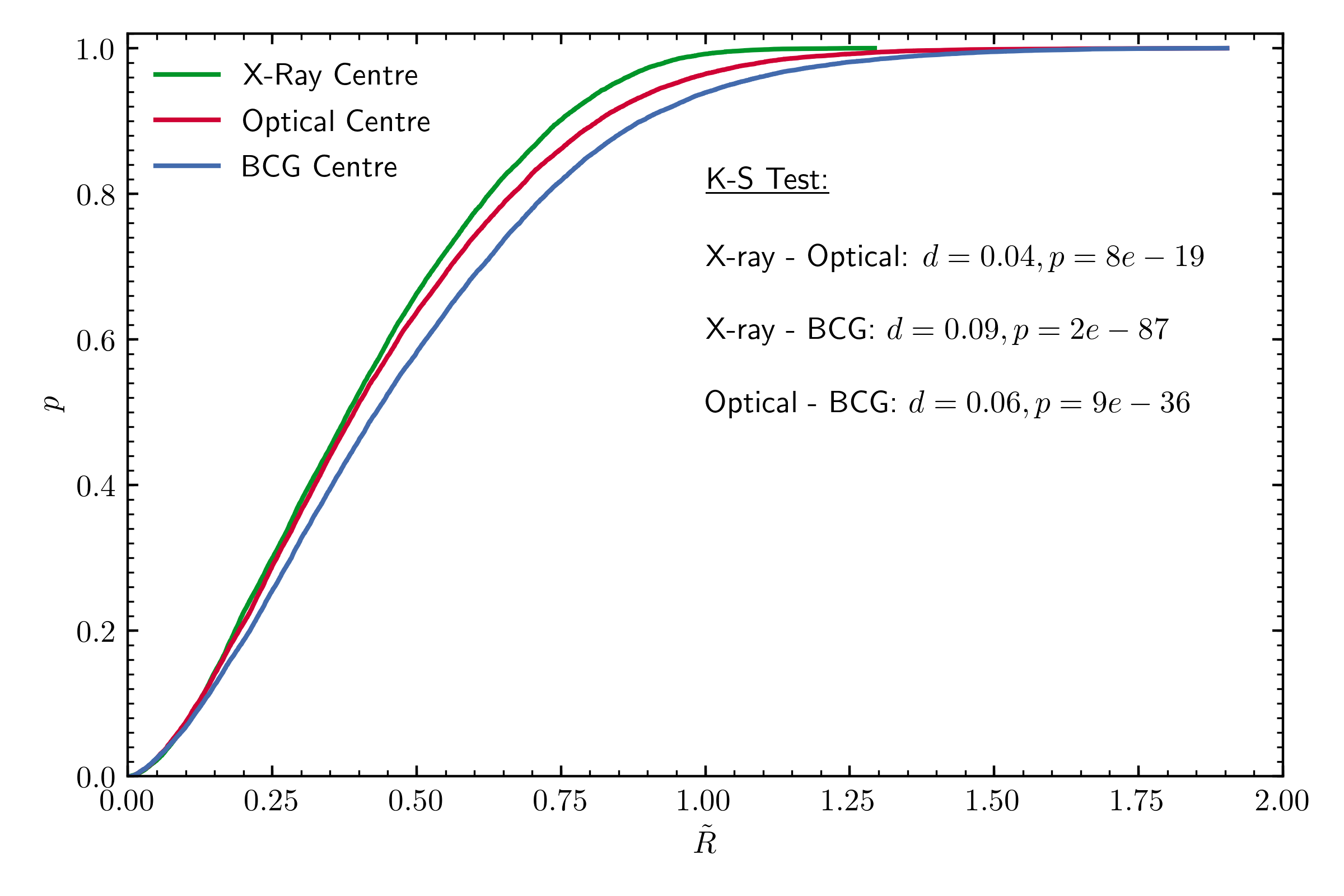}
    \caption{Cumulative distribution functions for all the galaxies used in each centroid analysis as a function of their projected separation from the cluster centre. A K-S test has been performed to check whether there is a significant difference between the populations used in each case. An extremely small $p$-value was found in each combination, so the null hypothesis that these are members of the same distribution can be rejected.}
    \label{fig:cdfcen}
\end{figure}

\begin{figure}
    \centering
    \includegraphics[width=\columnwidth]{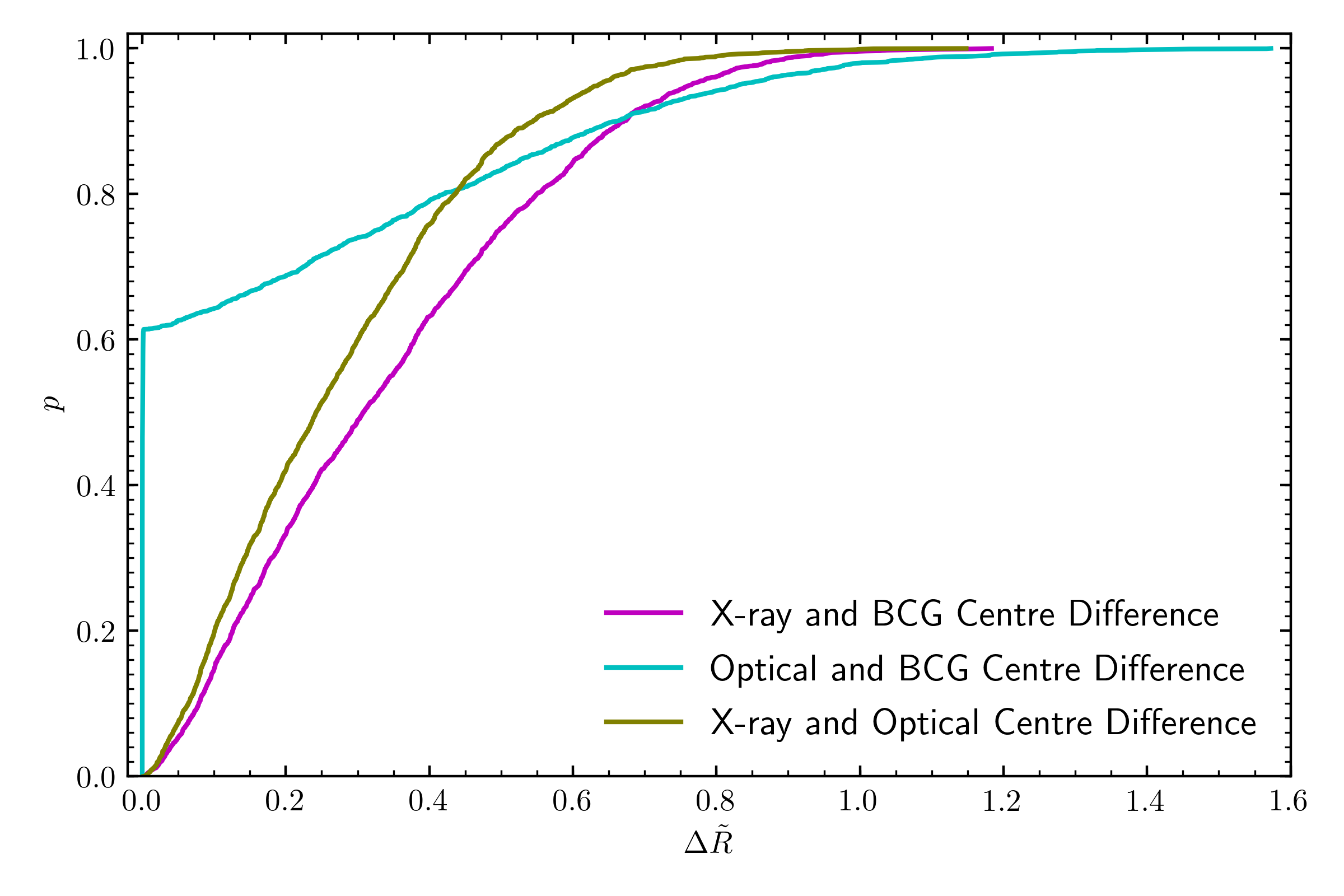}
    \caption{Cumulative distribution functions of the differences between the position of the centre for each centroid combination, for all clusters used in the analysis.}
    \label{fig:cendiff}
\end{figure}

To test whether each choice of centroid creates statistically distinct galaxy populations, a two sided Kolmogorov--Smirnov (K-S) test has been performed on the cumulative distribution functions (CDFs) of the positions of the galaxies used in each independent analysis (after the data reduction described in section \ref{sec:reduct} has been performed) from the cluster centre, shown in Fig. \ref{fig:cdfcen}. Each case is shown to represent a statistically distinct population. Fig. \ref{fig:cendiff} shows the CDFs of the difference of the central position for each centroid pair for clusters in SPIDERS, in units of the virial radius $r_{200}$. As expected for the optical and BCG centres there are many cases of coincidence ($61\%$ of the cluster population). 

These centre offsets can be informative in their own right. For example, the BCG-X-ray centre offset could be a probe on cluster substructure, and the dynamical state of a cluster \citep{BCGXRAYoffset}. The expectation is very small positional offsets for relaxed clusters, but non-negligible offsets for disturbed systems. The uncertainties of RASS X-ray positions translate to positional uncertainties ranging from $0.01-0.2 r_{200}$. Hence from Fig. \ref{fig:cendiff} it would appear that a significant fraction of these clusters are likely to be disturbed systems, consistent with \citet{PROPRIS}, \citet{XOffset} and references therein. Furthermore, \citet{XOffset} demonstrate that the X-ray centroid uncertainty tends to increase the observed X-ray-BCG/optical offset.


\subsection{Data reduction}

\label{sec:reduct}


\begin{table*}
\centering
\caption{Table of the different imposed conditions on all the SPIDERS DR16 galaxies and clusters, and the isolated effect of each. \textbf{X} = X-ray centroid, \textbf{O} = Optical centroid, \textbf{B} = BCG centroid.}
\begin{tabular}{@{}llll@{}}
\hline
SPIDERS parameter & Condition & Explanation & Isolated Effect \\ 
\hline
\texttt{NCOMPONENT} & $= 1$ &  No mergers or more than one cluster. & Removes 228 clusters. \\
\texttt{ALLZ\_NOQSO} & Use these redshifts & Removes Quasar templates. & Accurate spectroscopic redshifts. \\
\texttt{ALLZWARNING\_NOQSO} & $= 0$ &  Remove erroneous redshifts. & Removes 390 galaxies.\\
\texttt{ALLZ\_ERR\_NOQSO} & $< 0.0002$ & Remove galaxies with large redshift uncertainties. & Removes 3794 galaxies. \\
\texttt{R200C\_DEG} & $\tilde{R} \leq 2$ &  NFW model validity \citep{NFWrange}. & Removes 0 / 9 / 91 galaxies in \textbf{X} / \textbf{O} / \textbf{B}. \\
\texttt{SCREEN\_ISMEMBER\_W} & $= 1$ & Identified as a cluster member by all inspectors. & Removes 9\,716 galaxies. \\
\texttt{SCREEN\_CLUVDISP\_BEST} & $\lvert \Delta \rvert< 2.7\sigma_{\rm v}$ & Removes interlopers \citep{Mamon10,Mamon13}. & Removes 11\,839 / 8\,888 / 11\,406 galaxies in \textbf{X} / \textbf{O} / \textbf{B}. \\
\hline
\end{tabular}
\label{tab:conds}
\end{table*}

Table \ref{tab:conds} demonstrates the different conditions imposed on the raw SPIDERS DR16 dataset in an attempt to obtain an uncontaminated measurement of $\hat{\Delta}$. The effect of each condition, isolated from all the others, is also shown. The choice of a redshift uncertainty limit of $\sigma_{\rm z} < 0.0002$ includes $98.5\%$ of the population, and removes the upper tail of galaxies with large redshift uncertainties. There are also other, more subtle effects not included in the table. For the BCG and optical case $2512$ galaxies are automatically removed so the central galaxy of a cluster is not compared with itself, furthermore in some cases the BCG or CG has a redshift uncertainty $\geq 0.0002$, and so that cluster is not used in the analysis, since all the velocity offsets could potentially be biased. For the X-ray case, $5\,604$ galaxies within $\tilde{R} \leq 0.2$ are removed to find the averaged central redshift of the cluster. Finally, the restriction of there being a spectroscopically observed galaxy within  $\tilde{R} \leq 0.033$ of the optical centre also removes some clusters from the optical analysis, resulting in an extra $4\,405$ galaxies removed. The net effect of these conditions and those in Table \ref{tab:conds} is the removal of $22\,539$, $20\,563$ or $17\,052$ galaxies in the X-ray, optical or BCG centroid case respectively, from the original dataset which contains $41\,663$ spectroscopic redshifts. The remaining number of galaxies is greater than $10\,000$ in all cases, and so should yield a small enough standard error on $\hat{\Delta}$ to resolve it from zero effect.

High resolution X-ray data are needed to determine whether the large centroid offsets in Fig. \ref{fig:cendiff} are caused by clusters being in a disturbed state, something which could in principle bias the gravitational redshift signal. As this information is not available for SPIDERS DR16 clusters, no selection cut has been made based on the size of these centroid offsets. Tests were however performed to ensure that the removal of those clusters with the largest offsets had no significant effect on the results.

\begin{figure}
    \centering
    \includegraphics[width=0.65\linewidth]{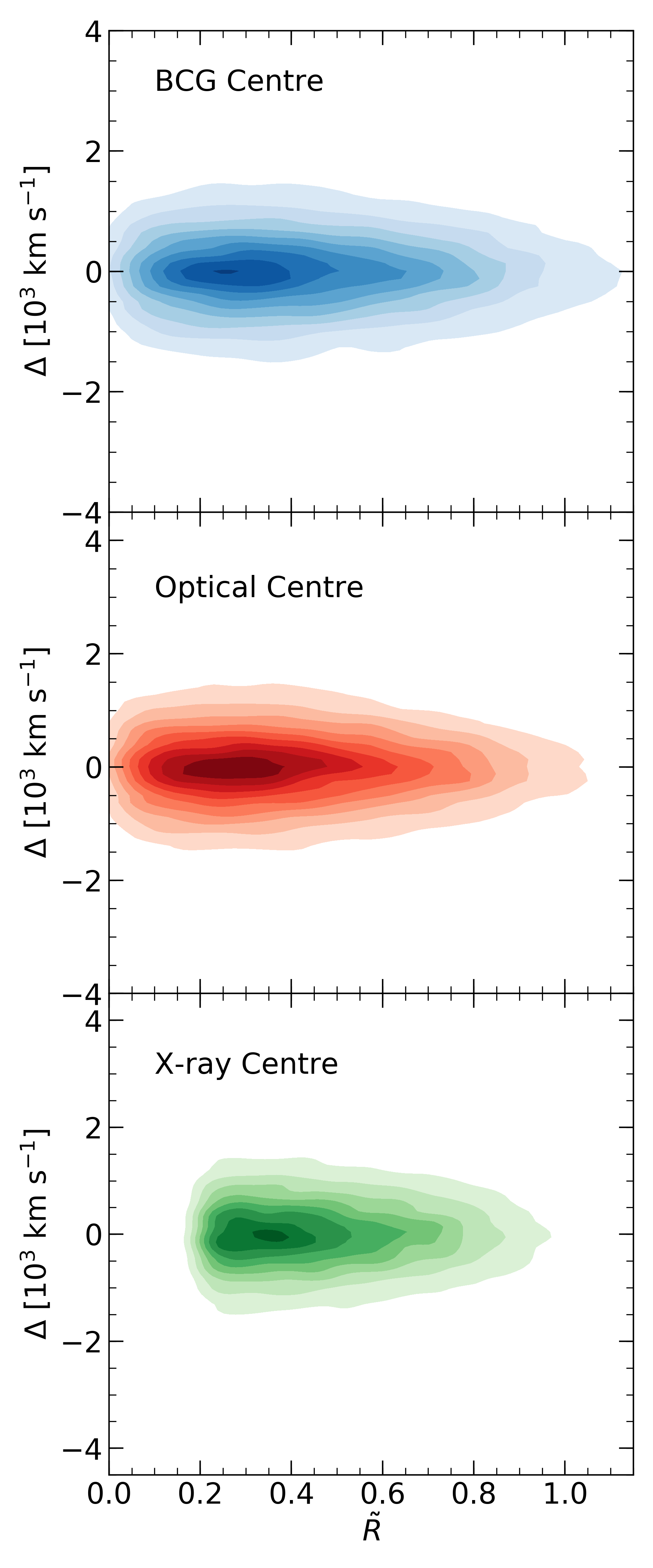}
    \caption{Density maps of the velocity offsets of all the remaining galaxies after refining the original SPIDERS catalogue, as a function of their projected distance from the cluster centre in units of the virial radius. In the BCG centre case (top) there are $24\,611$ galaxies, there are $21\,100$ for the optical (middle) and $19\,124$ for the X-ray (bottom).}
    \label{fig:drdv}
\end{figure}

Fig. \ref{fig:drdv} shows density maps of the remaining galaxies used in the final analysis for each centroid case. On the $x$-axis is their projected distance from the cluster centre in units of the virial radius, while on the $y$-axis is the size of each galaxy's velocity offset from the centre of its parent cluster.

\section{Contributions to \texorpdfstring{$\hat{\Delta}$}{TEXT}}
\label{sec:delta}
\subsection{Gravitational redshifting}

A distribution of velocity offsets between galaxies and their host cluster's centre is expected to have an average value that is blueshifted, as light experiences the largest redshifting at the minimum of the clusters potential well. For a single galaxy, the gravitational redshift, expressed as a velocity offset, is given by the difference between the gravitational potential at the galaxies distance from the cluster centre (here the dimensionless distance in units of $r_{200}$ is used), and that at the centre,\begin{equation}
    \Delta_{\rm gz}(\tilde{r}) = (\Phi(0) -\Phi(\tilde{r}))/c \, ,
\end{equation}
where the gravitational potential is that which is associated with an NFW dark matter density profile  (more detail is given in Appendix \ref{app:conc}). Only line-of-sight information can be measured, and therefore only the projected distance from the centre of the cluster, $\tilde{R}$, is known; see (\ref{eq:thetaR}). The density along the line of sight to that distance must be integrated along with the potential difference. Hence, for a single cluster \citep{NFWclus},
\begin{equation}
    \Delta_{\rm c,gz}(\tilde{R}) = \frac{2r_{200}}{c\Sigma(\tilde{R})}\int^{\infty}_{\tilde{R}}\left(\Phi(0)-\Phi(\tilde{r})\right)\frac{\rho(\tilde{r})\tilde{r}d\tilde{r}}{\sqrt{\tilde{r}^{2}-\tilde{R}^{2}}} \, ,
    \label{eq:DELTA}
\end{equation}
where $\Sigma(\tilde{R})$ is the surface mass density profile found from integrating the NFW density profile along the line of sight,
\begin{equation}
    \Sigma(\tilde{R}) = 2r_{200}\int_{\tilde{R}}^{\infty} \frac{\tilde{r}\rho(\tilde{r})}{\sqrt{\tilde{r}^{2}-\tilde{R}^{2}}} d\tilde{r} \, .
\end{equation}
By integrating with respect to $\tilde{r}$, and not the vector $\mathbf{\tilde{r}}$, spherical symmetry of the clusters is being assumed. Although often not the case for a single cluster, a stacked set of many clusters is expected to exhibit spherical symmetry.

Following \citet{WHH} the gravitational redshift signal for a stacked cluster sample can be calculated using
\begin{equation}
    \Delta_{gz} (\tilde{R}) = \frac{\int_{M_{\rm min}}^{M_{\rm max}} \Delta_{\rm c,gz}(\tilde{R}) \Sigma(\tilde{R})\left(dN/dM_{200}\right)dM_{200}}{\int_{M_{\rm min}}^{M_{\rm max}} \Sigma(\tilde{R})\left(dN/dM_{200}\right)dM_{200}} \, ,
    \label{eq:GRint}
\end{equation}
where the gravitational redshift profile for a single cluster has been convolved with the cluster mass distribution to accurately represent the stacked signal. 

\subsection{Transverse doppler effect}
\label{sec:TD}
The peculiar redshift of a galaxy can be decomposed as follows:
\begin{align}
    &\mbox{\boldmath$\beta$} = \frac{\mathbf{v}}{c}\,,\\
    &1+z_{\rm \rm \rm pec} \simeq 1 + \beta_{\rm los} + \beta^{2}/2 + \cdots \, ,
\end{align}
where $\beta_{\rm \rm los}$ gives the component in the line of sight. And so there is a second-order term due to transverse motion of the galaxy. This gives rise to the transverse doppler effect, which will contribute a small positive shift in the location of a velocity distribution; this is typically $\sim$few km\,s$^{-1}$, and is relatively constant with distance from the cluster centre.

To find the size of this effect for a set of galaxy velocity offsets from their host cluster's centre, we calculate
\begin{equation}
   \Delta_{\rm TD} = \left\langle v_{\rm gal}^{2} -v_{0}^{2} \right\rangle /2c \, .
\end{equation}
Calculating this effect involves a similar integral over the line-of-sight density profile and a convolution with the mass distribution \citep{Pea}. For a single cluster,
\begin{equation}
        \Delta_{\rm c,TD}(\tilde{R}) = \frac{2Qr_{200}}{c\Sigma(\tilde{R})}\int^{\infty}_{\tilde{R}}\left(\tilde{r}^{2}-\tilde{R}^{2}\right)\frac{d\Phi(\tilde{r})}{d\tilde{r}}\frac{\rho(\tilde{r})d\tilde{r}}{\sqrt{\tilde{r}^{2}-\tilde{R}^{2}}} \, ,
\end{equation}
where $Q = 3/2$ for isotropic orbits. This must be convolved with the mass distribution as in (\ref{eq:GRint}),
\begin{equation}
    \Delta_{\rm TD} (\tilde{R}) = \frac{\int_{M_{\rm min}}^{M_{\rm \rm max}} \Delta_{\rm c,TD}(\tilde{R}) \Sigma(\tilde{R})\left(dN/dM_{200}\right)dM_{200}}{\int_{M_{\rm min}}^{M_{\rm max}} \Sigma(\tilde{R})\left(dN/dM_{200}\right)dM_{200}} \, .
    \label{eq:TDint}
\end{equation}

\subsection{Light-cone effect}

The Universe is not static. Observations of galaxies lie in our past light cone, and as such there is some discrepancy between the distance observed between two sources, and the true distance. In between light being emitted from both sources, the second emitter will have moved a distance depending on its line-of-sight velocity. The relation between the separation expressed in light-cone coordinates and rest-frame coordinates is \citep{Kaiser}:
\begin{equation}
    dx_{\rm \rm LC} = \frac{dx_{\rm RF}}{1-v_{\rm x}/c}\, .
\end{equation}
This extra factor of $1/(1-v_{\rm x}/c)$ in the distance leads to an extra factor of $(1-v_{\rm x}/c)$ in the number density as $\rho \propto 1/V$ and for the cylindrical volume observed $V \propto dx$. This bias on the observed density of objects, dependent on their line-of-sight velocity, creates a bias on $\hat{\Delta}$.

Integrating over the line-of-sight coordinate $x$ gives a contribution proportional to $v_{\rm x}^{2}$:
\begin{equation}
  \Delta_{\rm LC} = \left\langle v_{\rm x, gal}^{2} -v_{\rm x,0}^{2}\right\rangle /c\, .
\end{equation}
Once again this gives a small positive contribution to the shifting of the location, opposite in sign to the effect of gravitational redshifting. Assuming isotropic orbits of the galaxies, we obtain
\begin{equation}
    \Delta_{\rm LC} = \frac{2}{3} \Delta_{\rm TD} \, .
\end{equation}

\subsection{Surface brightness modulation}
\label{sec:SB}
Galaxies in spectroscopic samples are chosen according to their apparent luminosity/magnitude. Due to the special relativistic beaming effect, this apparent luminosity can be changed by the peculiar motion of the galaxies. For galaxies lying just below the required apparent luminosity, motion towards the observer could shift them inside the cut, while those moving away could be shifted just outside the cut. Generally, this creates a small preferential bias in favour of galaxies moving towards the observer, with the overall effect of a small blueshifting on the centre of a distribution of velocity offsets.

The size of this effect depends strongly on the galaxy survey, for example in \citet{WHH} the flux limit is an $r$-band magnitude of $r=17.77$, while in SPIDERS the limit is an $i$-band fibre magnitude of $i=21.2$ in a $2''$ aperture \citep{Clerc}. To calculate the size of this effect, consider the fractional change in the apparent luminosity as a function of the spectral index at the cosmological redshift of the source, as well as the peculiar velocity of the source galaxy \citep{Kaiser}, given by
\begin{equation}
    \Delta L/L = \left(3 + \alpha(z_{\rm cos})\right)\frac{v_{\rm x}}{c} \, .
\end{equation}
The modulation of the number density of detectable objects is given by
\begin{equation}
    (\Delta L/L)\delta(z) = -\left(3 + \alpha(z)\right)\frac{v_{\rm x}}{c}\frac{d \ln{n}(>L_{\rm lim}(z))}{d \ln{L}} \, .
\end{equation}
Where $\delta(z)$ is the redshift dependent logarithmic derivative of the number distribution of galaxies. The redshift dependence comes from translating the apparent luminosity limit to an absolute luminosity limit that varies with redshift. Following \citet{Kaiser}, $\alpha(z)\simeq 2$ is assumed over the whole redshift range. Hence assuming isotropy:
\begin{align}
    \Delta_{\rm SB} &= -5\left\langle \delta(z)\right\rangle\left\langle v_{\rm \rm x}^{2} -v_{\rm x,0}^{2}\right\rangle /c\, ,\\
    \Delta_{\rm SB} &= -\frac{10}{3}\left\langle \delta(z)\right\rangle \Delta_{\rm TD} \, .
\end{align}
As $\left\langle \delta(z)\right\rangle$ is $\mathcal{O}(1)$, the shift due to the surface brightness modulation is the largest correction besides gravitational redshifting, and again is fairly constant with distance from the cluster centre.

An exact expression involves an investigation into the variation of $\delta(z)$ over the redshift range of the SPIDERS cluster catalogue, which we develop in section \ref{sec:predict}. 

\subsection{Combined effect}
\label{sec:deltacomb}
These effects are not the only ones present. \citet{stackedGR} give a comprehensive summary of the different contributions to $\hat{\Delta}$, including cross-terms. These are shown to reduce the $\hat{\Delta}$ signal by only $\lesssim 1$\,km\,s$^{-1}$, and so for the purposes of this analysis will not be considered further. 

Importantly, the Transverse Doppler effect discussed in section \ref{sec:TD} assumes that differences in redshift are being used to approximate velocity differences. The choice of using logarithmic differences alters the size of the Transverse Doppler effect. Assuming isotropy:
\begin{align}
    &\ln(1+z_{\rm pec}) \simeq \beta_{\rm x} + \beta_{\rm x}^{2} + \cdots\, ,\\
   &\Delta_{\rm TD*} =\left\langle v_{\rm x,gal}^{2} -v_{\rm x,0}^{2}\right\rangle /c \, , \\
   &\Delta_{\rm TD*} = \frac{2}{3}\Delta_{\rm TD}\,.   
\end{align}
The combination of the effects considered in this analysis gives
\begin{align}
   \hat{\Delta} &= \Delta_{\rm gz} + \Delta_{\rm LC} + \Delta_{\rm SB} + \Delta_{\rm TD*}\, , \\
   \hat{\Delta} &= \Delta_{\rm gz} + \left(2-5\left\langle \delta(z)\right\rangle\right)\frac{2}{3}\Delta_{\rm TD}\, . \label{eq:deltacomb}
\end{align}
And so from assuming isotropy only the gravitational and transverse doppler effects need to be calculated, as well as the logarithmic derivative $\delta(z)$, to account for the four largest contributions.

\section{Predicted \texorpdfstring{$\hat{\Delta}(\tilde{R})$}{TEXT}}
\label{sec:predict}
\begin{figure}
    \centering
    \includegraphics[width=\linewidth]{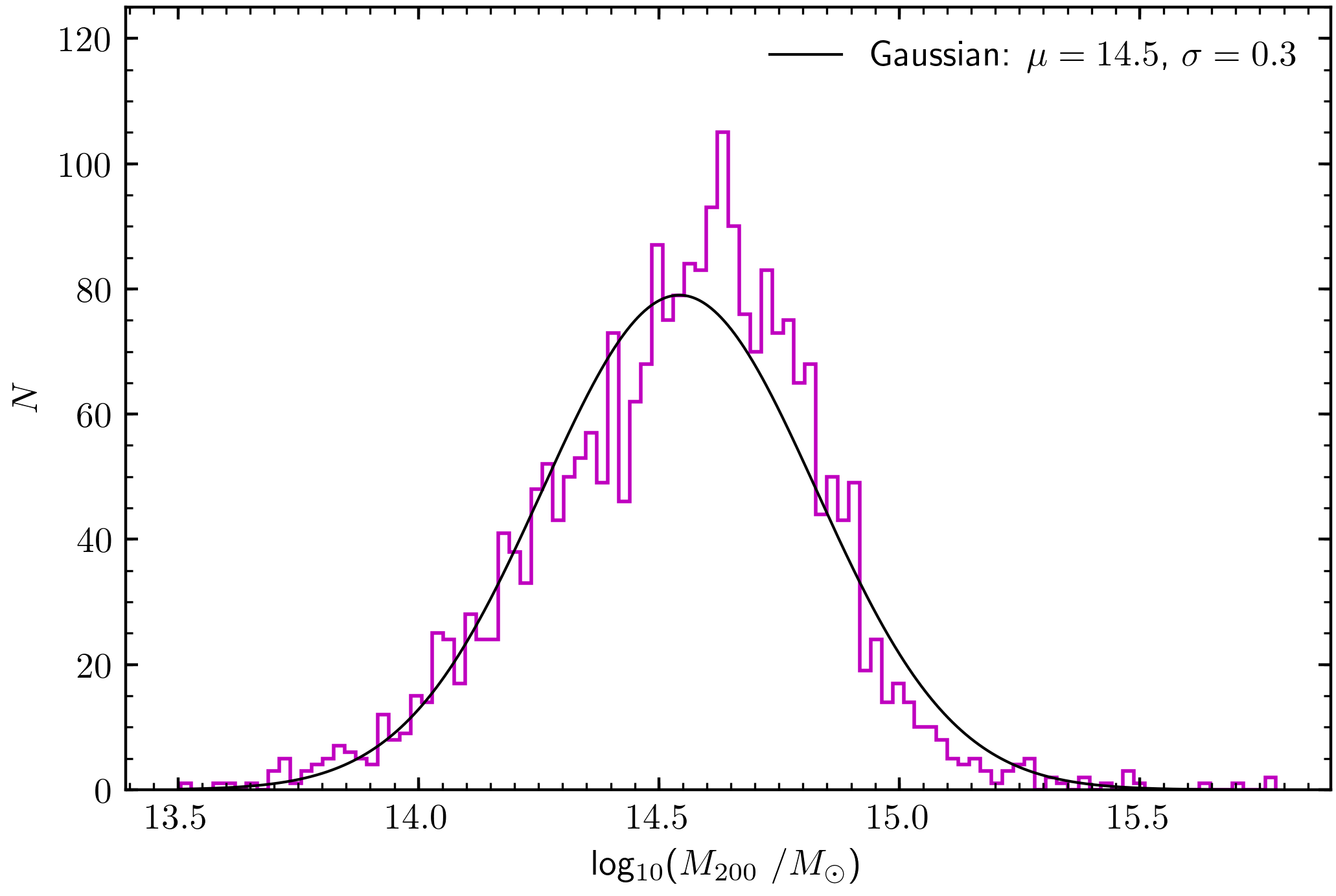}
    \caption{Distribution of the logarithm of virial masses of the SPIDERS clusters used in this analysis. The distribution can be approximately fitted with a Gaussian.}
    \label{fig:Mdist}
\end{figure}

To predict the size of $\hat{\Delta}$ for the stacked SPIDERS clusters, first an expression for the mass distribution is needed for the convolution in (\ref{eq:GRint}). An advantage of such a well studied cluster sample is there is pre-existing knowledge on the mass distribution, found from X-ray properties of the clusters. In \citet{WHH} the mass distribution needed to be estimated directly from the velocity distribution. The mass distribution of SPIDERS clusters used in the final analysis is shown in Fig. \ref{fig:Mdist}. It is approximated as a Gaussian with a mean of log$_{10}$($M_{200}/M_{\odot}$) $= 14.5$ and standard deviation of $0.3$. A skewed normal would provide a better fit, but the effect on the final result would be marginal.

\begin{figure}
    \centering
    \includegraphics[width=\linewidth]{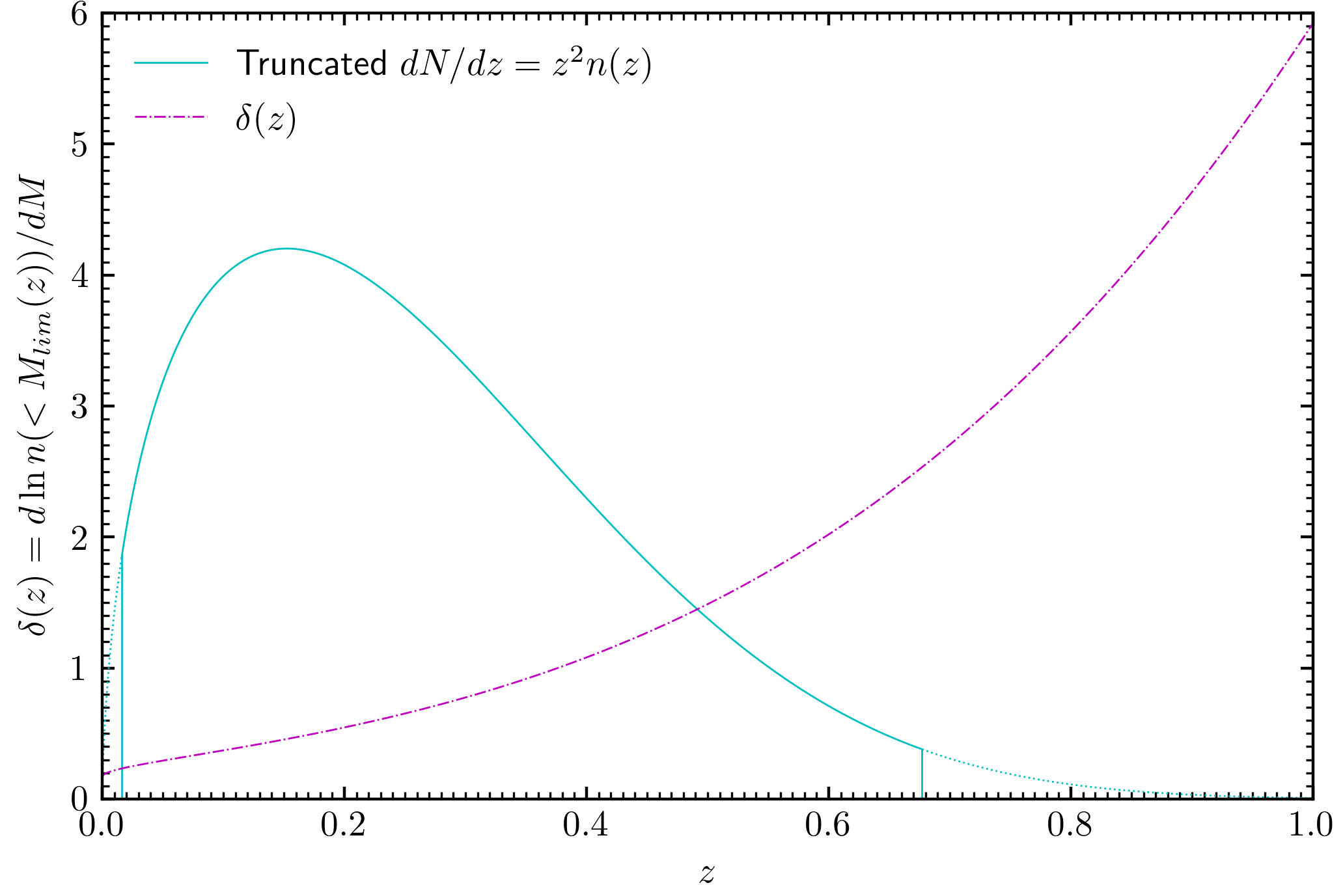}
    \caption{The blue-green solid curve gives the galaxy number distribution $dN/dz = z^{2} n(z)$, truncated at the host cluster maximum and minimum redshifts in SPIDERS. The purple dot dash cure shows the logarithmic derivative of the number density with respect to magnitude, limited by the absolute magnitude sensitivity, which is a function of redshift. Its average value over the relevant redshift range, restricted by $dN/dz = z^{2} n(z)$, is $\left\langle \delta(z)\right\rangle \simeq 0.7$.}
    \label{fig:SB}
\end{figure}


 
 \begin{figure}
     \centering
     \includegraphics[width=\linewidth]{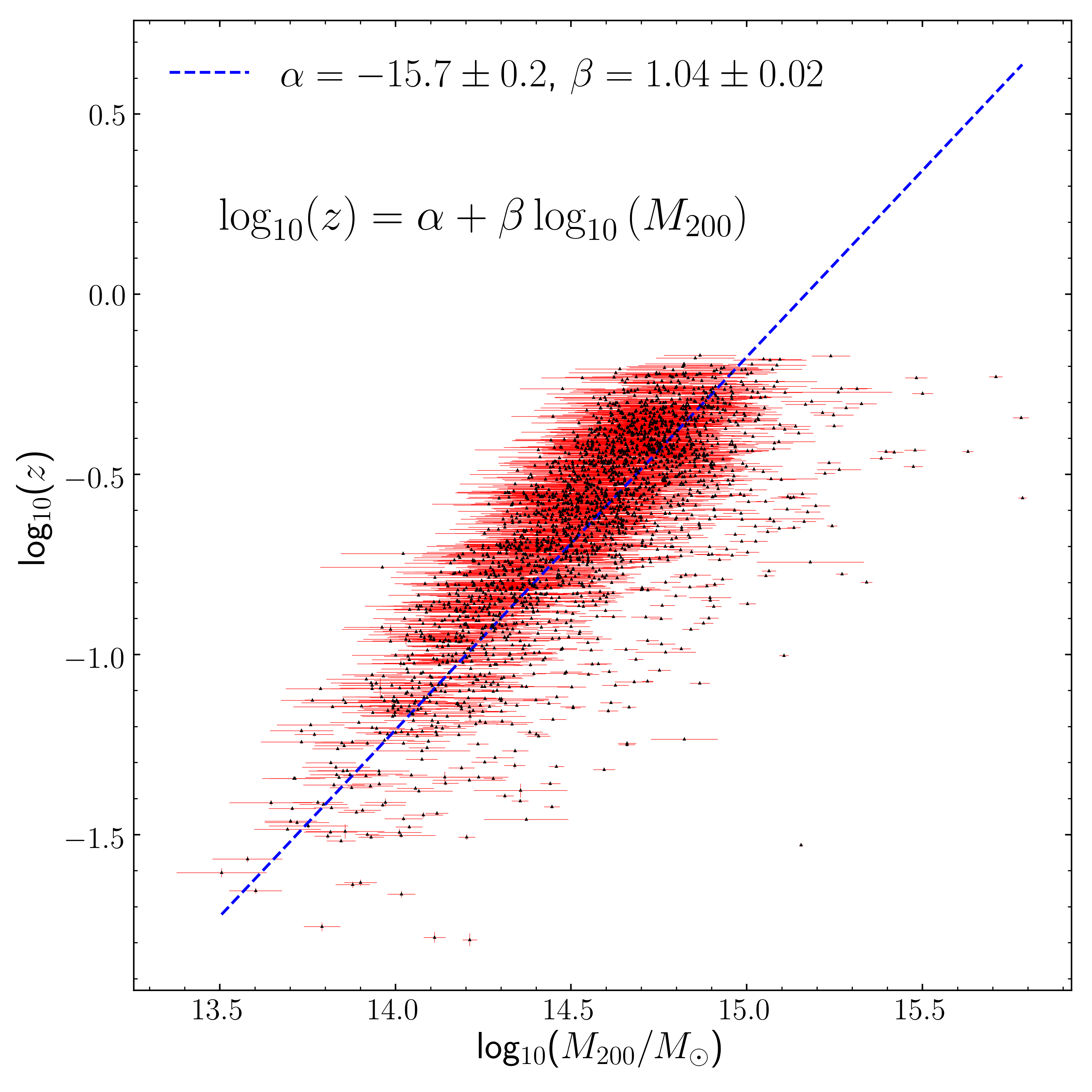}
     \caption{A redshift-mass relation for the clusters in the SPIDERS catalogue. The line of best fit is found using the BCES bisector method.}
     \label{fig:Mz}
 \end{figure}

The next step towards a prediction of $\hat{\Delta}$ requires the calculation of $\langle\delta(z)\rangle$ in equation (\ref{eq:deltacomb}). To find the surface brightness modulation in SPIDERS clusters the process demonstrated in \citet{Kaiser} has been closely followed, using values from \citet{LF}; the result is shown in Fig. \ref{fig:SB}. Two functions have been calculated: the logarithmic derivative $\delta(z)$ described in section \ref{sec:SB}, and the redshift dependent number distribution $dN/dz = z^{2}n(z)$. The fibre magnitude limit of $21.2$ has been used for the magnitude cut, and while galaxies at low redshift may be up to $\sim\!1$ mag brighter, this is found to only slightly increase the gradient of $\delta(z)$ at low redshift and hence have a minor impact on the result. The $y$-axis in Fig. \ref{fig:SB} shows the value of $\delta(z)$, and the number distribution has been scaled up for comparison. To find the average value of $\delta(z)$ over the redshift range of the cluster catalogue, restricted by the number distribution,
\begin{equation}
   \left\langle \delta(z)\right\rangle = \frac{\int\limits_{z_{\rm l}}^{z_{\rm u}} z^{2}n(z)\delta(z) \, dz}{\int\limits_{z_{\rm l}}^{z_{\rm u}} z^{2}n(z) \, dz} \, ,
\end{equation}
where $z_{\rm l} = 0.016$ and $z_{\rm u} = 0.667$ corresponding to the lower and upper redshift limit in the SPIDERS cluster catalogue. The result of this integration is
\begin{equation}
   \left\langle \delta(z)\right\rangle \simeq 0.7 \, .
    \label{eq:SBdelta}
\end{equation}

Finally, reasonable values for the concentration parameter $c_{200} = r_{200}/r_{\rm s}$, which gives the ratio between the virial radius and the so-called `scale radius' $r_{\rm s}$ are needed. This relates to the form of the NFW density profile, and the explicit dependence can be seen in Appendix \ref{app:conc}. In \citet{concmass} a redshift dependent $M_{200}-c_{200}$ relation is found:
\begin{multline}
    \log_{10}(c_{200}) = 0.52+(0.905-0.52)e^{-0.617z^{1.21}}\\
    - (0.101-0.026z)\log_{10}\left(\frac{M_{200}}{10^{12}h^{-1}M_{\odot}}\right) \, .
    \label{eq:c200}
\end{multline}
This replaces $c_{200}$ in the integration's in (\ref{eq:GRint}) and (\ref{eq:TDint}). A $z-M_{200}$ relation for the SPIDERS clusters, seen in Fig. \ref{fig:Mz}, is used to replace $z$ with $M_{200}$ in (\ref{eq:c200}). There is of course uncertainty in this relation—and the errors from the BCES bisector fit will be propagated through to the predictions for General Relativity and $f(R)$ gravity.

\subsection{General relativity}

\begin{figure}
    \centering
    \includegraphics[width=\linewidth]{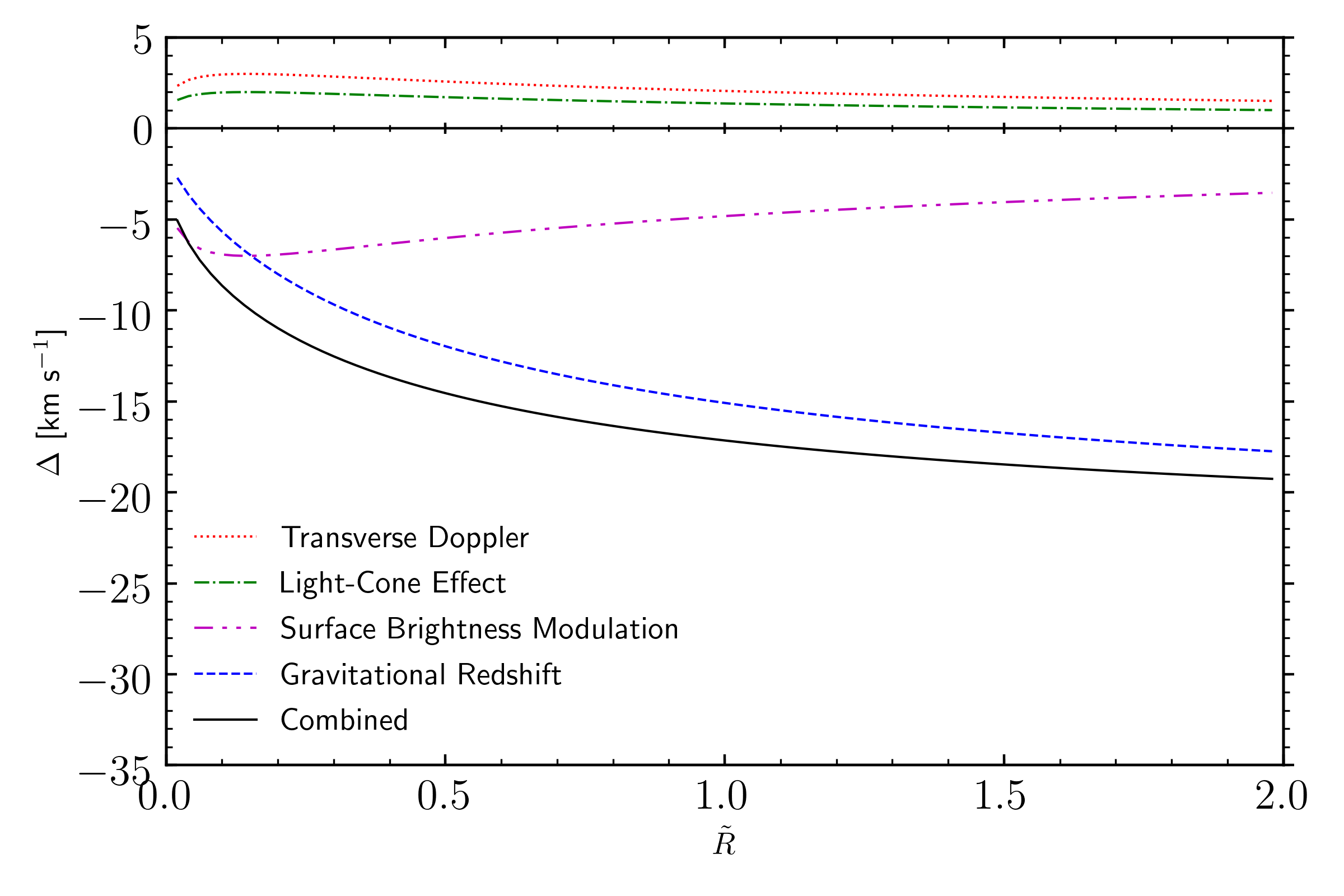}
    \caption{Different contributions to a General Relativity predicted shift of the average of galaxy velocities in the stacked SPIDERS clusters as a function of the projected distance from the centre of the cluster, in units of the virial radius $r_{200}$.}
    \label{fig:GR}
\end{figure}

Using (\ref{eq:SBdelta}), the combination of the effects in section \ref{sec:deltacomb} gives
\begin{equation}
    \hat{\Delta}_{\rm GR} = \Delta_{\rm gz} -\Delta_{\rm TD} \, .
\end{equation}
The results using fiducial values are demonstrated in Fig. \ref{fig:GR}. The net effect of the other contributions is a small added blueshifting to the gravitational redshift.

\subsection{\texorpdfstring{$f(R)$}{TEXT} gravity}

For General Relativity with a cosmological constant—the standard model of cosmology $\Lambda$CDM—the Einstein-Hilbert action, which is integrated over all coordinates of spacetime, describes the interaction between matter and gravity:
\begin{equation}
\label{eq:einhil}
       S = \int d^{4}x \, \sqrt{-g}\left[ \frac{M^{2}_{\text{pl}}}{2} \left( R-2 \Lambda \right) +\mathcal{L}_{\rm m}\right] \, .
\end{equation}
Here, \smash{$M_{\rm pl}^{2} = 1 / 8\pi G$} is the reduced Planck mass; $R$ is the Ricci scalar, which gives information on the curvature of spacetime; $\Lambda$ is the cosmological constant; $\mathcal{L}_{\rm m}$ is the matter Lagrangian and $g$ is the determinant of the Friedmann-Lemaître-Robertson-Walker (FLRW) metric describing a homogeneous and isotropic expanding Universe.

A simple modification can be made to this action, representing a simple modification to gravity \citep{fR}:
\begin{equation}
\label{eq:fR}
       S = \int d^{4}x \, \sqrt{-g}\left[ \frac{M^{2}_{\rm pl}}{2} \left( R + f(R) \right) +\mathcal{L}_{\rm m}\right] \, ,
\end{equation}
where the cosmological constant $\Lambda$ has been replaced by some unknown function of the Ricci scalar.  In \citet{FAB} it is shown that, in the limit where the background value of $f(R)$ is much larger than a cluster's potential well, the effect on the gravitational force experienced by a test mass in the cluster due to this modification to gravity is $G \to 4/3 \,G$, and in the reverse scenario there is no modification to $G$. A strong field model with $|f_{\rm R0}| = 10^{-4}$ is shown to cause the $4/3$ enhancement for all halo masses used in their simulations, and this is the condition that will be assumed in this study for a simple comparison between the prediction of $\hat{\Delta}$ in General Relativity and $f(R)$ gravity. It should be noted that constraints on $f(R)$ gravity \citep[e.g.][]{fRcon} rule out this universal $4/3$ enhancement for all cluster masses, but it is emphasised that this simple model is used to provide some insight into the sensitivity of $\hat{\Delta}$ to variations on the theory of gravity.

\section{Results}
\label{sec:res}
\begin{figure}
    \centering
    \includegraphics[width=\columnwidth]{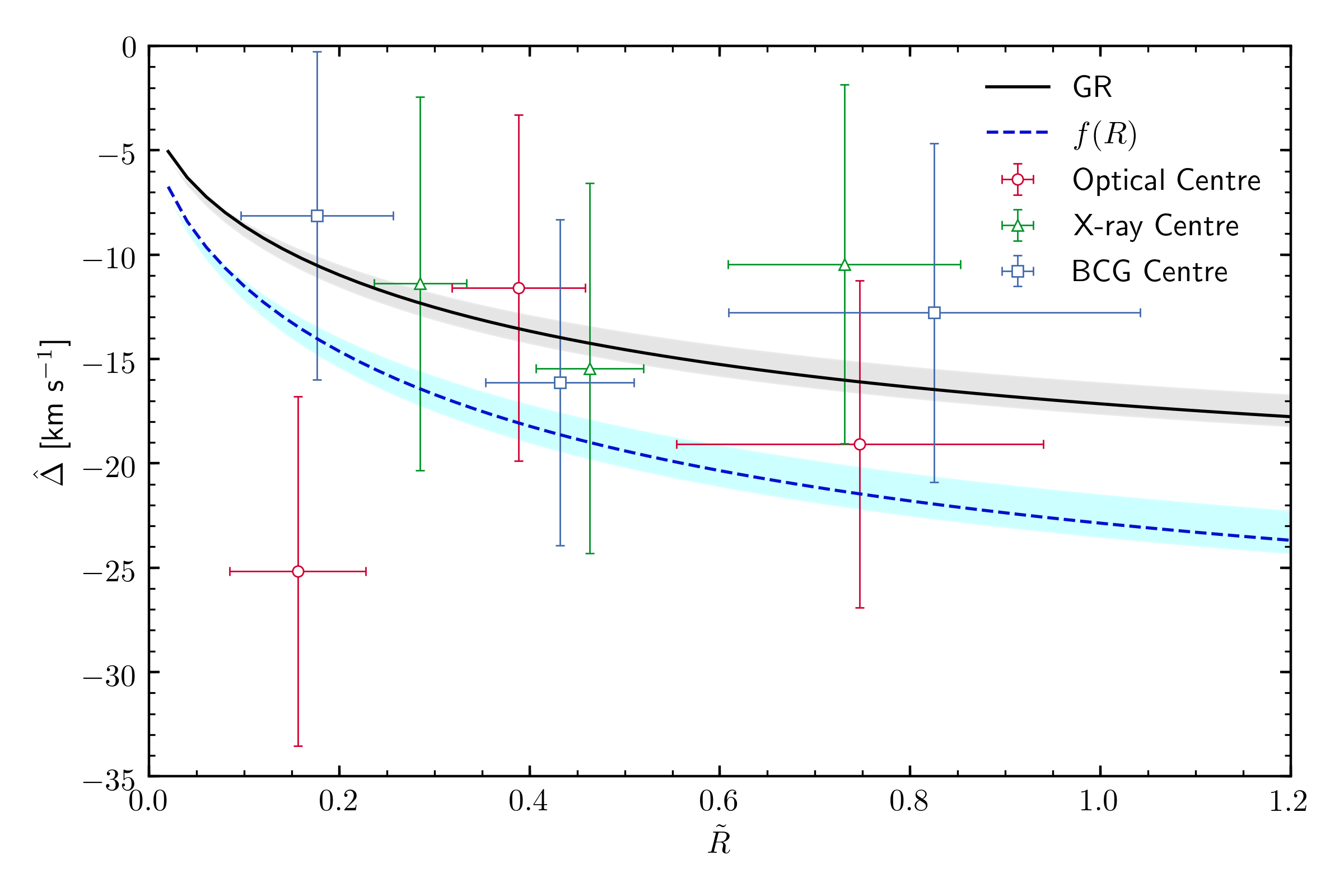}
    \caption{The biweight location of the distribution of SPIDERS galaxy velocity offsets using  three definitions of the cluster centre, as a function of their projected distance from the centre in units of the virial radius $r_{200}$. Equal numbers of galaxies have been used for each bin. The observations are compared with predictions for $\hat{\Delta}$ from General Relativity (GR) and $f(R)$ gravity. The highlighted regions around these predictions demonstrate error bounds propagated through from the redshift-mass relation.}
    \label{fig:gz}
\end{figure}

In Fig. \ref{fig:gz} the galaxy velocity offsets for each centroid, shown in Fig. \ref{fig:drdv}, have been split into three bins with equal numbers. The biweight average of the distribution of $\Delta$'s in each bin is found, giving the $y$-value of each data point, while its value on the $x$-axis is the average projected distance of the binned galaxies from the cluster centre in units of the virial radius $r_{200}$. It is the projected distance as our observations only measure angles on the sky, so there is some ignorance as to how far away from the cluster centre each galaxy truly lies. The $y$-error gives the standard error of the average value, while the $x$-error gives the dispersion of galaxy positions within each bin. All errors give regions of $68\%$ probability. Three bins have been chosen to maximise the number of galaxies per bin, while still allowing easy visual comparison with the predictions of two theories of gravity: General Relativity (solid black line) and $f(R)$ gravity (blue dashed line). The highlighted region around each prediction shows error bounds caused by the uncertainty in the redshift-mass relation in Fig. \ref{fig:Mz}.

In the BCG and the X-ray case there is good agreement with the General Relativity predicted variation of $\hat{\Delta}$ with $\tilde{R}$. The apparent tension between the GR prediction and the first optical data point is not significant, and is lessened when the data are rebinned. Dependence of the optical result on the allowed difference between a cluster's optical centroid and the nearest spectroscopically observed galaxy was also tested. All results agreed to within $1\sigma$.

\begin{figure}
    \centering
    \includegraphics[width=\linewidth]{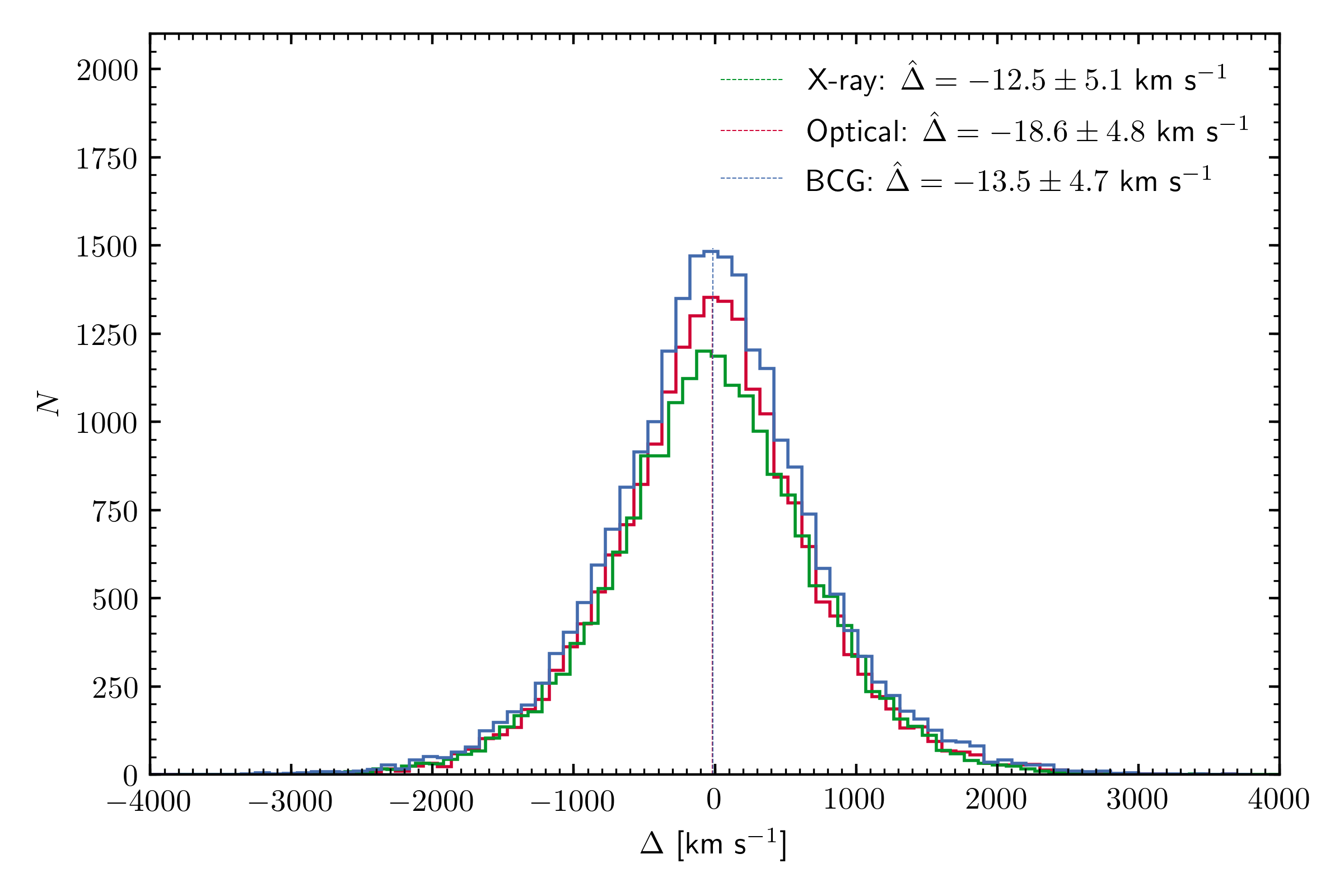}
    \caption{The velocity distributions and the integrated $\hat{\Delta}$ shift over the range $0 < \tilde{R} \leq 1$ for the stacked set of galaxies in each centroid case. Each bin has a width of $100$\,km\,s$^{-1}$.}
    \label{fig:int}
\end{figure}

Another informative way of presenting these results is through the total integrated $\hat{\Delta}$ over a defined range of $\tilde{R}$. These results are shown in Fig. \ref{fig:int}, alongside the distribution of galaxies used for each centroid to find the integrated effect. This further highlights the difference in numbers for each case. Only galaxies in the range $0<\tilde{R}\leq 1$ have been used as the number past this distance drops off rapidly, and there needs to be a specific distance range to compare the integrated effect of each centroid to the predictions of General Relativity and $f(R)$ gravity.

The integrated General Relativity and $f(R)$ signals in the range $0<\tilde{R}\leq 1$ are
\begin{align}
    &\hat{\Delta}_{\rm GR} = -13.7^{+0.7}_{-0.6} \text{\,km\,s}^{-1}\,,\\
    &\hat{\Delta}_{\rm f(R)} = -18.3^{+1.0}_{-0.8} \text{\,km\,s}^{-1}\,.
\end{align}

Hence all centroid cases are consistent with both GR and $f(R)$ to $1\sigma$, while the BCG and X-ray centre cases show more similarity to the GR prediction. The optical integrated effect appears to be more consistent with the $f(R)$ prediction, but in Fig. \ref{fig:gz} the evolution of $\hat{\Delta}$ in the optical case does not particularly follow that of $f(R)$, while the BCG and X-ray cases do have similar evolution to GR. Further, the BCG, X-ray and optical cases show a $\sim\!2.9\sigma$, $\sim\!2.5\sigma$ and $\sim\!3.9\sigma$ clear detection of $\hat{\Delta}$ respectively in the range $0<\tilde{R}\leq 1$.

These results are broadly consistent with others in the literature. In \citet{Kaiser}, an updated prediction for the stacked clusters in \citet{WHH}, using the combination of effects described in section \ref{sec:deltacomb} (and other small contributions) gives $\hat{\Delta}_{\rm GR} = -11.8$\,km\,s$^{-1}$. Also the observational result in \citet{WHH} is $\hat{\Delta} = -7.7 \pm 3.0$\,km\,s$^{-1}$. Considering they quote a mean mass of $\sim\!2\times 10^{14} M_{\odot}$, and the distribution in Fig. \ref{fig:Mdist} peaks around $3.2 \times 10^{14} M_{\odot}$, and is skewed towards higher masses, the GR prediction for the SPIDERS clusters and the size of the observed $\hat{\Delta}$ seem to follow the expected behaviour of a higher mass sample leading a larger predicted $\hat{\Delta}$. Furthermore, \citet{GRSDSSBOSS} found $\hat{\Delta} = -11^{+7}_{-5}$\,km\,s$^{-1}$ using SDSS Run 10 galaxies and clusters. This result is also in good agreement, albeit with large uncertainties.

\section{Discussion and Conclusions}
\label{sec:conc}

A positive detection of the gravitational redshift effect, along with other small contributions to a shift of the average of a distribution of galaxy velocity offsets, denoted $\hat{\Delta}$, is reported using SPIDERS DR16 galaxies and clusters. This work considered three definitions of the centre of a cluster: using the Brightest Cluster Galaxy; though a probabilistic determination of a red-sequence Central Galaxy; or from using the peak of X-ray emission. Each definition provides a distinct galaxy population, and produces results for $\hat{\Delta}$ largely consistent with one another. Most notably the X-ray and BCG centroid cases predict a very similar change of $\hat{\Delta}$ with projected distance from the cluster centre, $\tilde{R}$. This is despite the need for a slightly cumbersome definition of the central redshift in the X-ray case.

Galaxy redshift errors have not been used when finding $\hat\Delta$ as they are likely to be correlated with the apparent magnitude and galaxy type, which could introduce a bias on $\hat\Delta$. However it is important to note that the uncertainty in observed redshifts could still introduce a bias, yet it is hoped the large numbers of galaxies used beats down this systematic.

The result with the smallest error (largest sample of galaxies and clusters) and most robust methods comes from using a BCG to trace the centre of a cluster. Using the centre of X-ray emission to trace the cluster centre is a promising method: it removes the issue of accidental miscentring on foreground or background galaxies, and in dynamic clusters where the BCG is unlikely to trace the centre of mass, X-ray centres may be a more accurate measure. The downside in this analysis was the large X-ray centroid uncertainty in faint ROSAT sources and a cumbersome central redshift definition—from finding the average redshift of galaxies in the core region. In general, a combination of these two methods—using the BCG closest to the X-ray centre, could provide a powerful hybrid, combining X-ray's lack of contamination and the ease of observing a BCG.

For the optical case, although redMaPPer assigns an optical centre based on a most likely Central Galaxy, in many cases there was no spectroscopically observed galaxy near to the optical centre. For some clusters this could simply be due to positional errors, but in others it is likely that the CG identified by redMaPPer has not been spectroscopically observed by SPIDERS. Despite this shortcoming, finding the CG using a probabilistic approach has potential benefits over simply using a BCG. In cases where the cluster is highly dynamic, the filters used by redMaPPer may identify a CG more appropriately.

The integrated results for $\hat{\Delta}$ in the range $0<\tilde{R}\leq1$ found in each centroid case were consistent with both theories of gravity—General Relativity and $f(R)$ gravity—to within $1\sigma$; however,  $\hat{\Delta}(\tilde{R}$) slightly favours General Relativity in the BCG and X-ray cases. Each centroid case demonstrates a significant ($>2.5\sigma$) detection of the gravitational redshifting of galaxies in SPIDERS clusters.

Possible improvements to this work include a more robust prediction for the size of $\hat{\Delta}$ in the theories of gravity used, involving better treatment of the redshift dependence, and a skewed normal fit to the mass distribution. Furthermore, comparison with the predicted value of $\hat{\Delta}$ in other theories of gravity than the two considered could reveal the usefulness of this approach. If most other alternative theories have very similar predictions of $\hat{\Delta}(\tilde{R}$) to GR, then because it is such a small effect with often large uncertainties, the efficacy of the method may be limited.

SDSS-V using \textit{eROSITA} X-ray data \citep{SDSSV} promises more galaxy clusters with lower masses up to larger redshifts. More clusters means a better constrained $\hat{\Delta}$, and better prospects for using this signal to distinguish between theories of gravity. \textit{eROSITA} will also have much better X-ray resolution, giving more localised X-ray central positions. The 4MOST \textit{eROSITA} Galaxy Cluster Redshift Survey \citep{4MOST} aims to provide spectroscopic redshifts for $\sim\!40\,000$ \textit{eROSITA} galaxy groups/clusters, including their BCG and >15 cluster members for $z<0.7$. By combining an X-ray central position found from \textit{eROSITA} data and the nearest 4MOST BCG or redMaPPer identified CG to this X-ray position, there is the potential for accurate identification of cluster centres, even in dynamic systems where simply using a BCG causes miscentring.

To obtain a very strong positive detection of $\hat{\Delta}$, say $>10\sigma$, consider the BCG case with an uncertainty of $\pm 1$\,km\,s$^{-1}$. Assuming a similar velocity dispersion, there needs to be $\sim\!530\,000$ galaxies in the whole sample. While this is around an order of magnitude larger than what is currently possible with SPIDERS, with forthcoming deep optical telescopes such as \textit{The Vera C. Rubin Observatory} for galaxy identification and, for example, the \textit{Euclid} satellite for spectroscopic follow-up, this is certainly an achievable goal. \textit{The Vera C. Rubin Observatory} will overall observe billions of galaxies \citep{LSST}, and \textit{Euclid}'s Near Infrared Spectrometer plans to measure $\sim\!50$ million spectroscopic redshifts of galaxies \citep{EUCLID}. These numbers, coupled with well measured X-ray selected clusters from \textit{eROSITA}, promise tightly constrained measurements of $\hat{\Delta}$ in the near future.

The same error of $\sim\!1$\,km\,s$^{-1}$ for the BCG result in this analysis, compared with the $f(R)$ prediction, would indicate a $\sim4\sigma$ deviation between observations and the prediction of this example of an alternative theory of gravity.

Although these considerations demonstrate the sensitivity of  gravity theories to gravitational redshifting using galaxy clusters, an important caveat is that kinematic data alone is insufficient to provide adequate discrimination between theories of gravity. There is a degeneracy between the size of $G$ affecting the velocity distribution and the mass of the cluster—both GR and $f(R)$ can give rise to the same gravitational redshift signal but with different dark matter halo functions \citep{Pea}. Knowledge on how the X-ray inferred cluster mass changes for a given $f(R)$ is needed \citep{fRXray1,fRXray2}. Furthermore, as weak lensing based cluster mass estimates are unaffected by an extension to $f(R)$ \citep{WLmasses,WLfR}, this is a potential method by which this degeneracy can be broken.

\section*{Acknowledgements}

We thank the referee for helpful suggestions leading to tests that improved the robustness of our results. CTM and CAC acknowledge support from Liverpool John Moores University. JAP was supported by the European Research Council under grant no. 670193. AS is supported by the ERC-StG `ClustersXCosmo’ grant agreement 716762, and by the FARE-MIUR grant 'ClustersXEuclid' R165SBKTMA.

\section*{Data Availability}

SPIDERS data are available through membership of the SDSS-IV UK Participation Group funded by LJMU.

Funding for the Sloan Digital Sky Survey IV has been provided by the Alfred P. Sloan Foundation, the U.S. Department of Energy Office of Science, and the Participating Institutions. SDSS acknowledges support and resources from the Center for High-Performance Computing at the University of Utah. The SDSS web site is www.sdss.org.

SDSS is managed by the Astrophysical Research Consortium for the Participating Institutions of the SDSS Collaboration including the Brazilian Participation Group, the Carnegie Institution for Science, Carnegie Mellon University, Center for Astrophysics | Harvard \& Smithsonian (CfA), the Chilean Participation Group, the French Participation Group, Instituto de Astrofísica de Canarias, The Johns Hopkins University, Kavli Institute for the Physics and Mathematics of the Universe (IPMU) / University of Tokyo, the Korean Participation Group, Lawrence Berkeley National Laboratory, Leibniz Institut f{\"u}r Astrophysik Potsdam (AIP), Max-Planck-Institut für Astronomie (MPIA Heidelberg), Max-Planck-Institut f{\"u}r Astrophysik (MPA Garching), Max-Planck-Institut f{\"u}r Extraterrestrische Physik (MPE), National Astronomical Observatories of China, New Mexico State University, New York University, University of Notre Dame, Observat{\'o}rio Nacional / MCTI, The Ohio State University, Pennsylvania State University, Shanghai Astronomical Observatory, United Kingdom Participation Group, Universidad Nacional Aut{\`o}noma de M{\'e}xico, University of Arizona, University of Colorado Boulder, University of Oxford, University of Portsmouth, University of Utah, University of Virginia, University of Washington, University of Wisconsin, Vanderbilt University, and Yale University.



\bibliographystyle{mnras}
\bibliography{GZ}




\appendix

\section{The NFW Density Profile}
\label{app:conc}
The NFW density profile gives the mass density as a function of the distance from the centre of a cluster in units of its virial radius \citep{NFWclus}:\begin{align}
    \rho(\tilde{r}) &= \frac{M_{200}c_{200}^{2}g(c_{200})}{4\pi r_{200}^{3}\tilde{r}(1+c_{200}\tilde{r})^{2}} \, ,\\
    \Phi(\tilde{r}) &= -g(c_{200})\frac{GM_{200}}{r_{200}}\frac{\ln{(1+c_{200}\tilde{r})}}{\tilde{r}} \, .
\end{align}
Important parameters here are the concentration parameter $c_{200}$ and the function $g(c_{200})$:
\begin{align}
    &c_{200} = \frac{r_{200}}{r_{s}} \, ,\\
    &g(c_{200}) = \left(\ln{(1+c_{200})}-c_{200}/(1+c_{200})\right)^{-1} \, .
\end{align}
The concentration parameter gives the ratio between the virial radius of an astronomical body and its so called scale radius, and gives an indication of the mass concentration of the object. Typically, for clusters $c_{200} \sim 5$ and for bright galaxies $c_{200} \sim 10$ .

\section{Bootstrapping Test}
\label{app:bs}

\begin{figure}
    \centering
    \includegraphics[width=0.65\linewidth]{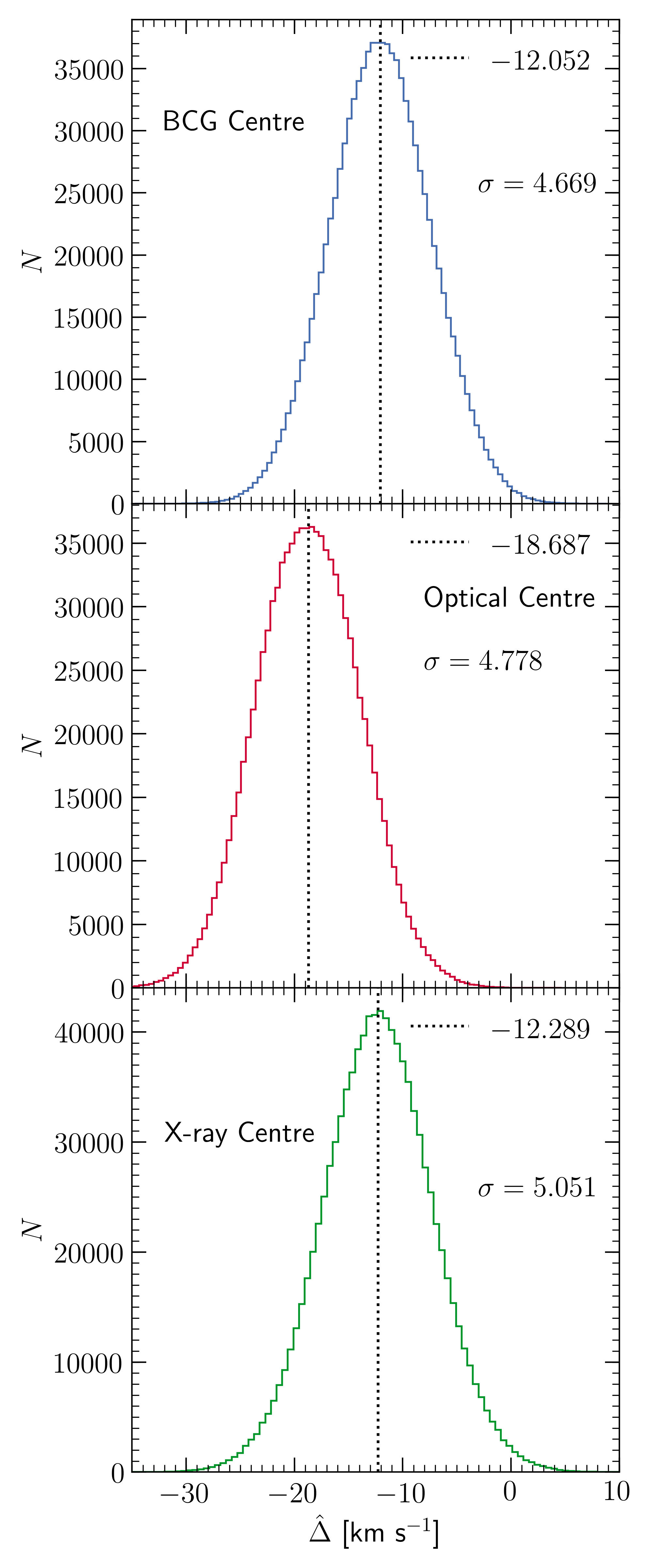}
    \caption{Bootstrapping with $10^{6}$ samples on the galaxy velocity offset distributions for each centroid case.}
    \label{fig:bs}
\end{figure}

Sampling with replacement was used to obtain $10^{6}$ values for $\hat\Delta$ for the galaxy samples in each of the centroid cases to test the consistency of the quoted uncertainties in Fig. \ref{fig:gz}; histograms of the results are shown in Fig. \ref{fig:bs}. The average values and standard deviations of the resulting Gaussians are consistent with the quoted results.

\bsp	
\label{lastpage}
\end{document}